\documentclass{aa}
\usepackage{graphicx}
\graphicspath{{figures/}}
\usepackage{newtxtext, newtxmath}
\usepackage[locale=US]{siunitx}
\DeclareSymbolFont{upgreek}{U}{eur}{m}{n}
\DeclareMathSymbol{\umu}{0}{upgreek}{"16}
\DeclareSIPrefix{\micro}{\text{\ensuremath{\umu}}}{-6}
\usepackage{xcolor}
\definecolor{darkblue}{rgb}{0,0,0.55}
\definecolor{darkmagenta}{rgb}{0.55,0,0.55}
\usepackage[colorlinks=true, allcolors=darkblue]{hyperref}
\usepackage{orcidlink}
\begin{document}

    \title{Scattered polarized radiation of extrasolar circumplanetary rings}
    
    
    
    \author{
        M. Lietzow\,\orcidlink{0000-0001-9511-3371}\and
        S. Wolf\,\orcidlink{0000-0001-7841-3452}
        }

    \authorrunning{M. Lietzow\and S. Wolf}
    
    \institute{
        Institute of Theoretical Physics and Astrophysics,
        Kiel University, Leibnizstr. 15, 24118 Kiel, Germany\\
        \email{mlietzow@astrophysik.uni-kiel.de}
        }
    
    \date{Received / Accepted}


    \abstract
    {}
    {We have investigated the impact of circumplanetary rings consisting of spherical micrometer-sized particles on the net scattered light polarization of extrasolar gas giants.}
    {Using the three-dimensional Monte Carlo radiative transfer code POLARIS, we studied the impact of the macroscopic parameters that define the ring, such as its radius and inclination, and the chemical composition of the ring particles on the net scattered polarization.
    For the spherical ring particles, we applied the Mie scattering theory.
    We studied the flux and polarization of the scattered stellar radiation as a function of planetary phase angle and wavelength from the optical to the near-infrared.}
    {For the chosen grain size distribution, the dust particles in the ring show strong forward scattering at the considered wavelengths.
    Thus, the reflected flux of the planet dominates the total reflected and polarized flux at small phase angles.
    However, the scattered and polarized flux of the ring increase at large phase angles and exceeds the total reflected planetary flux.
    For large rings that contain silicate particles, the total reflected flux is dominated by the radiation scattered by the dust in the ring at all phase angles.
    As a result, the orientation of polarization is parallel to the scattering plane at small phase angles.
    In contrast, for a ring that contains water ice particles, the orientation of polarization is parallel to the scattering plane at large phase angles.
    Depending on the ring inclination and orientation, the total reflected and polarized flux show a specific distribution as well.
    Large particles show a strong polarization at large phase angles compared to smaller particles.
    For a Jupiter-like atmosphere that contains methane and aerosols, methane absorption features are missing in the spectrum of a ringed planet.}
    {Scattering of the stellar radiation by dust in circumplanetary rings of extrasolar planets results in unique features in the phase-angle- and wavelength-dependent reflected and polarized net flux.
    Thus, exoplanet polarimetry provides the means to study not only the planetary atmosphere and surface, but also to identify the existence and constrain the properties of exoplanetary rings.}


    \maketitle
%

\section{Introduction}
\label{sec:introduction}

In addition to the prominent ring system of Saturn, rings have also been discovered around Jupiter \citep{smith-etal-1979}, Uranus \citep{elliot-etal-1977, smith-etal-1986}, and Neptune \citep{hubbard-etal-1986, smith-etal-1989}.
Therefore, it is reasonable to assume that extrasolar planets may show the same phenomena.
The detection and characterization of circumplanetary rings of extrasolar planets potentially provides insights into the formation of planets and rings and the obliquity of the planet \citep{schlichting-chang-2011}, as well as into the formation of satellites and extrasolar moons \citep{mamajek-etal-2012}.
However, these so-called exorings have not been detected for any of the over \num{5000} confirmed extrasolar planets (as of November 2022;
\href{http://exoplanet.eu/}{The Extrasolar Planets Encyclopaedia}, \citealt{schneider-etal-2011};
\href{https://exoplanetarchive.ipac.caltech.edu/}{The NASA Exoplanet Archive}, \citealt{akeson-etal-2013}).

Nevertheless, several authors have investigated the detectability  of rings of transiting extrasolar planets and predicted their photometric and spectroscopic signatures \citep[e.g.,][]{barnes-fortney-2004, ohta-etal-2009, tusnski-valio-2011, zuluaga-etal-2015, sucerquia-etal-2017, akinsanmi-etal-2018}.
In addition, observations of light curves have been compared with ring models, such as for 51~Peg~b, but in this case the detected reflected light in principle cannot be explained by a ring system \citep{santos-etal-2015}.
However, ring-like structures are consistent with observations and have been proposed for the systems J1407 \citep{mamajek-etal-2012} and PDS~110 \citep{osborn-etal-2017}.
Furthermore, \citet{piro-vissapragada-2020} explored whether super-puffs, which are planets with very low mean densities due to their large radii of $\sim$\SIrange{4}{10}{R_\oplus} but relatively low masses of $\sim$\SIrange{2}{6}{M_\oplus} \citep{lee-chiang-2016, wang-dai-2019, libby-roberts-etal-2020}, can be explained as ringed exoplanets.
Finally, the scattered and reflected light for a ringed extrasolar planet has specific photometric signatures, making it detectable even for non-transiting systems \citep{arnold-schneider-2004, dyudina-etal-2005, sucerquia-etal-2020, zuluaga-etal-2022}.

However, previous approaches to investigating reflected light have usually assumed simplified models, such as diffuse reflecting or transmitting rings, or planets approximated by Lambertian spheres.
In addition, multiple scattering, a precise scattering phase function based on the ring particles, and polarization due to scattering were neglected.
The radiation reflected by planets is usually polarized with an amplitude on the order of \SI{10}{ppm} for unresolved hot Jupiters \citep{seager-etal-2000}.
In particular, in the case of HD~189733, \citet{bott-etal-2016} reported an amplitude of about \SI{29}{ppm} for the polarization variations.
For the Sun, the net polarization is measured to be below \SI{10}{ppm} \citep{kemp-etal-1987}.
For nearby FGK dwarfs, the polarization is on the order of \SI{10}{ppm} \citep{bailey-etal-2010, cotton-etal-2017}.
However, for active stars, the polarization can be on the order of about \SIrange{20}{40}{ppm} \citep{cotton-etal-2017}, while for M dwarfs the polarization reaches values of about \SI{100}{ppm} \citep{bailey-etal-2010} and is, thus, significantly larger than the planetary contribution.

While polarimetry is used in our Solar System to characterize planetary atmospheres \citep{dollfus-coffeen-1970, dollfus-1975, tomasko-smith-1982, west-hart-hord-etal-1983, west-hart-simmons-etal-1983, hansen-hovenier-1974, smith-tomasko-1984, tomasko-doose-1984, schmid-etal-2011, mclean-etal-2017} and circumplanetary rings \citep{kemp-murphy-1973, hall-riley-1974, dollfus-1979, esposito-etal-1980, johnson-etal-1980}, it is also proposed for the search for Earth-like extrasolar planets \citep{bailey-2007, stam-2008, karalidi-etal-2011, sterzik-etal-2012, song-qu-2017}, to detect and characterize Jupiter-like extrasolar planets \citep{stam-etal-2004}, and to characterize the cloud compositions of extrasolar planets \citep{lietzow-wolf-2022}.
\citet{lietzow-etal-2021} simulated the total reflected and polarized flux of a gas giant with a circumplanetary ring, assuming multiple scattering, Rayleigh scattering by molecular hydrogen in the atmosphere, and Mie scattering by small water ice particles in the ring.
The authors find that the reflected polarized flux is also affected by an additional circumplanetary ring.
Thus, polarization measurements are potentially a useful tool for detecting and characterizing not only extrasolar planets, but also the planetary environment, such as circumplanetary rings.

The aim of this study is to investigate the impact of circumplanetary rings consisting of spherical micrometer-sized particles on the net scattered light polarization of extrasolar gas giants.
For this purpose, we studied the reflected flux and its polarization as a function of planetary phase angle at optical to near-infrared wavelengths.
In Sect.~\ref{sec:methods} we describe the numerical methods that are used to calculate the scattered polarized radiation.
In Sect.~\ref{sec:setup} we introduce our model setup, which includes the planetary atmosphere as well as the geometric and physical properties of the circumplanetary ring.
Our results are presented in Sect.~\ref{sec:results}.
We mainly focus on close-in gas giants in which Rayleigh scattering is dominating.
A ringed cloudy Jupiter-like planet is also considered in Sect.~\ref{subsec:results-cloudy}.
Finally, in Sect.~\ref{sec:conclusions} we summarize our study.

\section{Methods}
\label{sec:methods}

The scattered polarized radiation was calculated using the publicly available three-dimensional Monte Carlo radiative transfer code POLARIS%
\footnote{%
\url{https://portia.astrophysik.uni-kiel.de/polaris}}
\citep{reissl-etal-2016}.
Recently, it has been extended to calculate the radiation scattered in a planetary atmosphere and the planetary environment as well \citep{lietzow-etal-2021}.

To describe the state and degree of polarization of the radiation, we use the Stokes formalism \citep[e.g.,][]{bohren-huffman-1983}.
Here, the radiation is characterized by its total flux, $I$, the components $Q$ and $U$, which describe the linear polarization, and the component $V$ for the circular polarization.
These parameters define the vector $\boldsymbol{S} = (I, Q, U, V)^\mathrm{T}$.
It is multiplied by a scattering matrix (Müller matrix), $\mathbf{F}(\vartheta),$ at each scattering event to account for a change in polarization,
\begin{equation}
    \label{eq:stokes-scattering}
    \boldsymbol{S}' \propto \mathbf{F}(\vartheta) \cdot \boldsymbol{S},
\end{equation}
where $\boldsymbol{S}'$ is the Stokes vector after the scattering event and $\vartheta$ is the scattering angle.
The linear polarized intensity and the degree of linear polarization are given by
\begin{equation}
    PI = \sqrt{Q^2 + U^2},\quad P = \frac{\sqrt{Q^2 + U^2}}{I},
\end{equation}
respectively.
The orientation of linear polarization is defined as
\begin{equation}
    \label{eq:pol-angle}
    \tan(2\chi) = \frac{U}{Q}.
\end{equation}
For radiation that is polarized parallel to the scattering plane, that is, the plane containing the star, planet, and observer, it is $\chi = \ang{0}$.
The wavelength- and phase-angle-dependent flux reaching the observer, $I_\lambda$, is given in \si{W.m^{-2}.m^{-1}} and, unless stated otherwise, is normalized to the planetary flux at a planetary phase angle of $\alpha = \ang{0}$ at a wavelength of \SI{0.55}{\um}.
Here, the phase angle, $\alpha$, is the angle between observer-planet and planet-star; thus, at $\alpha = \ang{0}$, the planet is at full phase.
For single scattered radiation, it is $\alpha = \ang{180} - \vartheta$.


\section{Model setup}
\label{sec:setup}

\subsection{General parameters}
\label{subsec:general-parameters}

\begin{figure}
    \centering
    \includegraphics[width=8.8cm]{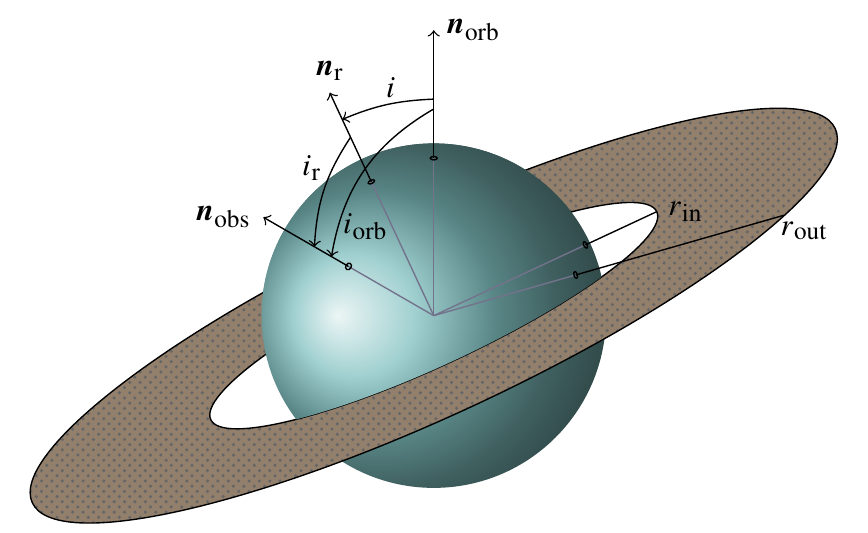}
    \caption{Illustration of the planet-ring model setup. See Sect.~\ref{sec:setup} for details.}
    \label{fig:setup}
\end{figure}

The planet with radius, $r_\mathrm{p}$, was illuminated by a Sun-like star with a spectrum that is approximated by Planck's law for a temperature of \SI{6000}{K}.
We assumed that the emitted stellar radiation is unpolarized.
The setup of the planet-ring model is shown in Fig.~\ref{fig:setup}.
The size of the ring is defined by its inner radius $r_\mathrm{in}$ and outer radius $r_\mathrm{out}$.
The vectors $\boldsymbol{n}_\mathrm{orb}$, $\boldsymbol{n}_\mathrm{r}$, and $\boldsymbol{n}_\mathrm{obs}$ define the normal of the planetary orbital plane, the normal of the ring plane, and the direction toward the observer, respectively.
The corresponding angles are given by
\begin{align*}
    &\cos i = \boldsymbol{n}_\mathrm{orb} \cdot \boldsymbol{n}_\mathrm{r},\\
    &\cos i_\mathrm{r} = \boldsymbol{n}_\mathrm{r} \cdot \boldsymbol{n}_\mathrm{obs},\\
    &\cos i_\mathrm{orb} = \boldsymbol{n}_\mathrm{orb} \cdot \boldsymbol{n}_\mathrm{obs}.
\end{align*}
Throughout this study, we assumed an observer who is located in the orbital plane, that is, $i_\mathrm{orb} = \ang{90}$.

\subsection{Atmospheric model}
\label{subsec:atmospheric-model}

We considered an atmosphere with properties dominated by gas.
The gaseous component consists of molecular hydrogen to account for Rayleigh scattering, as it is the most abundant gas in a gas giant.
The refractive index, the scattering cross section, and the scattering matrix of molecular hydrogen were calculated using the formulas given by \citet{keady-kilcrease-2000}, \citet{sneep-ubachs-2005}, and \citet{hansen-travis-1974}, respectively.
The pressure ranges from \SI{e-5}{bar} at the top of the atmosphere to \SI{100}{bar}, which defines the lower boundary.
Below this gaseous layer, we assumed a thick cloud layer that absorbs all incident radiation.
At a wavelength of \SI{0.55}{\um} and assuming a Jupiter-like gravity, the scattering optical depth of the gas is approximately \num{12}.

For this setup, the resulting geometric albedo is somewhat large, with a value of about \num{0.72} (see Sect.~\ref{subsec:results-general}).
For a semi-infinite Rayleigh scattering atmosphere, the geometric albedo is \num{0.7975} \citep{prather-1974}.
In contrast, hot gas giants ($\gtrsim$\SI{900}{K}) are expected to have low geometric albedos due to various opacity sources and missing reflecting clouds \citep{marley-etal-1999, sudarsky-etal-2000, sudarsky-etal-2005, cahoy-etal-2010, cahoy-etal-2017-erratum, cowan-agol-2011}.
In particular, sodium and potassium are very important opacity sources in the visible wavelength range, whereas methane and water are important in the near-infrared \citep{tsuji-etal-1999, burrows-etal-2000, sudarsky-etal-2000}.
For example, very low albedos ($<$\num{0.2}) were found for CoRoT-1b \citep{alonso-alapini-etal-2009, snellen-etal-2009}, CoRoT-2b \citep{alonso-guillot-etal-2009, snellen-etal-2010}, HAT-P-7b \citep{welsh-etal-2010}, HD~189733b \citep{evans-etal-2013}, HD~209458b \citep{rowe-etal-2006, rowe-etal-2008, brandeker-etal-2022}, and TrES-2b \citep{kipping-spiegel-2011}, as well as for some confirmed \textit{Kepler} giant planets \citep{angerhausen-etal-2015}.
For temperatures $\lesssim$\SI{350}{K}, water or ammonia condense and form reflecting clouds in the upper atmospheric layers, so the geometric albedo increases.
Therefore, gas giants in our Solar System have high geometric albedos in the visible wavelength range, such as \num{0.52} or \num{0.47} for Jupiter or Saturn \citep[e.g.,][]{tholen-etal-2000}, respectively.

Although cool Jupiters, which are gas giants with orbital periods longer than \SI{100}{d}, are much more common than close-in hot Jupiters \citep{wittenmyer-etal-2020}, we mainly focused on close-in gas giants since these planets are currently the best targets for the observation of reflected stellar radiation.
Instead of modeling a real planetary atmosphere, the simple non-absorbing Rayleigh scattering atmosphere served as a reference model to set an upper limit for the reflected flux.
Furthermore, the case of pure Rayleigh scattering is often used for radiative transfer calculations in planetary atmospheres and has been extensively studied \citep[e.g.,][]{chandrasekhar-1960, coulson-etal-1960, kattawar-adams-1971, prather-1974, van-de-hulst-1980, sromovsky-2005, buenzli-schmid-2009, natraj-etal-2009}.
It has also been reported that some hot Jupiters have atmospheres, in which the scattering is dominated by Rayleigh scattering of small particles, such as HD~189733b \citep{pont-etal-2008, pont-etal-2013}.
However, this is potentially due to a haze containing sub-micrometer-sized particles.
In addition to a purely scattering atmosphere, we reduced the single scattering albedo of the gaseous particles in Sect.~\ref{subsec:results-general} to investigate the case of a low geometric albedo of the planet as well.

In Sect.~\ref{subsec:results-cloudy} we considered a cool gas giant, similar to Saturn or Jupiter, for which additional cloud and haze particles are present in the atmosphere \citep[see, e.g.,][]{sato-hansen-1979, smith-1986, west-etal-1986, karkoschka-tomasko-1992, karkoschka-tomasko-1993, ortiz-etal-1996, west-etal-2004}.
For this purpose, we assumed small spherical particles with a standard size distribution \citep{hansen-1971},
\begin{equation}
    \label{eq:size-clouds}
    n(s) \propto s^{(1 - 3 \varv_\mathrm{eff}) / \varv_\mathrm{eff}}\ \exp[-s / (s_\mathrm{eff} \varv_\mathrm{eff})],
\end{equation}
where $s_\mathrm{eff}$, and $\varv_\mathrm{eff}$ are the mean effective radius and effective variance, respectively.
We used Jupiter-like atmospheric properties similar to the model atmosphere by \citet{mclean-etal-2017} that best fits their polarimetric observations of Jupiter.
The model considers an optically thin haze layer in the upper atmospheric layers above an optically thick cloud layer in the deeper atmosphere.
The haze layer is located between \SI{133}{mbar} and \SI{100}{mbar} with an optical depth of \num{0.2}.
For the haze particles, we assumed an effective radius of \SI{0.2}{\um} and an effective variance of \num{0.01}.
The refractive index is $\num{1.5} + \num{e-3} \mathrm{i}$ and represents the optical properties of benzene and polycyclic aromatic hydrocarbons, which are proposed to be present in the polar hazes \citep{friedson-etal-2002}.
The cloud layer has a top pressure of \SI{1}{bar} and an optical depth of \num{50}.
For the cloud particles, we assumed an effective radius of \SI{0.5}{\um} and an effective variance of \num{0.05}.
The cloud particles have a refractive index of $\num{1.42} + \num{1.5e-2} \mathrm{i}$ and represent the optical properties of an ammonia-like composition.
The scattering and absorption cross sections as well as the scattering matrix were calculated using the code \emph{miex} \citep{wolf-voshchinnikov-2004}, which is based on the Mie scattering theory \citep{mie-1908}.
The general model parameters are summarized in Table~\ref{tab:model-parameter}.

\begin{table*}
    \centering
    \caption{General model parameters.}
    \label{tab:model-parameter}
    \begin{tabular}{l l l l l}
        \hline\hline
        \noalign{\smallskip}
        & Gas & Clouds \tablefootmark{(b)} & Haze \tablefootmark{(b)} & Ring\\
        \noalign{\smallskip}
        \hline
        \noalign{\smallskip}
        Dimensions  & $\num{100} \geq p / \si{bar} \geq \num{e-5}$ & $\num{100} \geq p / \si{bar} \geq \num{1}$ & $\num{0.133} \geq p / \si{bar} \geq \num{0.1}$ & $\num{1.2} \leq r / r_\mathrm{p} \leq \num{2.3}$\\
        Composition & H$_2$, CH$_4$ \tablefootmark{(b)} & Ammonia-like & Benzene-like & Silicate mixture\\
        Size distribution & -- & Eq.~(\ref{eq:size-clouds}), & Eq.~(\ref{eq:size-clouds}), & Eq.~(\ref{eq:size-ring}),\\
        & -- & $s_\mathrm{eff} = \SI{0.5}{\um}$, & $s_\mathrm{eff} = \SI{0.2}{\um}$, & $\num{0.1} \leq s / \si{\um} \leq \num{100}$,\\
        & -- & $\varv_\mathrm{eff} = \num{0.05}$ & $\varv_\mathrm{eff} = \num{0.01}$ & $q = \num{-3}$\\
        Optical depth \tablefootmark{(a)} & $\sim$\num{12} (scattering) & \num{50} & \num{0.2} & \num{1} (vertical)\\
        \noalign{\smallskip}
        \hline
    \end{tabular}
    \tablefoot{
    \tablefoottext{a}{at \SI{0.55}{\um}.}
    \tablefoottext{b}{Methane, clouds, and a haze layer were used only in Sect.~\ref{subsec:results-cloudy}.}
    }
\end{table*}

\subsection{Ring model}
\label{subsec:ring-model}

For a star-planet separation below
\begin{equation}
    \label{eq:ice-distance}
    \left( \frac{L_\star}{16 \pi \sigma T_\mathrm{sub}^4} \right)^{1/2} \approx \num{2.7}\left( \frac{L_\star}{\si{L_\odot}} \right)^{1/2} \si{au},
\end{equation}
where $L_\star$ is the stellar luminosity, $\sigma$ is the Stefan-Boltzmann constant, and $T_\mathrm{sub} \approx \SI{170}{K}$ is the sublimation temperature of water ice, circumplanetary rings cannot be composed of water ice \citep{gaudi-etal-2003}.
Thus, circumplanetary rings of close-in planets are potentially composed of rocky materials, such as silicates or graphites.
This is consistent with the ring system of Saturn, where mainly icy material is assumed \citep[e.g.,][]{cook-etal-1973, cuzzi-etal-1984, cuzzi-etal-2009}.
For the Jovian ring, \citet{neugebauer-etal-1981} reported that its spectrum is consistent with reflection from rocky material.
Thus, unless stated otherwise, we assumed ring particles that are composed of a silicate mixture that represents the optical properties of interstellar dust \citep{draine-2003}.

The size distribution $n(s)$ of the ring particles between a minimum grain size $s_\mathrm{min}$ and a maximum grain size $s_\mathrm{max}$ is described by a simple power law with a power law index, $q$,
\begin{equation}
    \label{eq:size-ring}
    n(s) \propto s^q
\end{equation}
that represents the outcome of an ideal collisional cascade \citep{dohnanyi-1969}.
For the main ring of Saturn, the power law index is about \num{-3} \citep{cuzzi-pollack-1978}, while a power law index of about \num{-4.6} was found in Saturn's F ring \citep{showalter-etal-1992}.
For the rings of Jupiter, a power law index of about \num{-3.5} is consistent with observations \citep{tyler-etal-1981}.
For our default ring model, unless stated otherwise, we assumed small spherical micrometer-sized particles (\SI{0.1}{\um} $\leq s \leq$ \SI{100}{\um}, $q = \num{-3}$).
Micrometer-sized particles have also been proposed for the Jovian ring system \citep{showalter-etal-1987}, whereas the main ring system of Saturn mainly consists of centimeter and meter-sized particles \citep{cuzzi-pollack-1978, french-nicholson-2000}.

As in the case of the clouds and haze particles (see Sect.~\ref{subsec:atmospheric-model}), the scattering and absorption cross sections as well as the scattering matrix of the ring particles were calculated with the code \emph{miex} \citep{wolf-voshchinnikov-2004}.
Although dust particles in planetary rings are not spherical, optical properties of nonspherical particles can be approximated with equivalent spherical particles \citep{grenfell-warren-1999, neshyba-etal-2003}.
In contrast, the scattering phase function of large irregularly shaped particles differs from that of spheres \citep{asano-sato-1980, pollack-cuzzi-1980, mishchenko-travis-1994}.
One approach is to approximate the scattering by the Henyey-Greenstein function \citep{henyey-greenstein-1941}, as for example in the case of debris disks observations \citep[e.g.,][]{mccabe-etal-2002, schneider-etal-2006, schneider-etal-2009, stark-etal-2014, milli-etal-2017, lovell-etal-2022}.
\citet{hedman-stark-2015} reported that a linear combination of three Henyey-Greenstein functions is a good approximation for the particles inside the G and D ring of Saturn.
However, the phase curves of these rings were also fitted equally well using the Mie theory \citep{hedman-stark-2015}.
In addition, polarization characteristics such as the rainbow at large scattering angles are less pronounced or absent for nonspherical particles compared to spherical particles \citep{perry-etal-1978, asano-sato-1980, cai-liou-1982, takano-jayaweera-1985, hess-1998, goloub-etal-2000}.

The default ring extends from \num{1.2}$r_\mathrm{p}$ to \num{2.3}$r_\mathrm{p}$, similar to Saturn's main ring \citep[e.g.,][]{tholen-etal-2000}.
The ring has an opening angle of \ang{;;2}, resulting in a maximum vertical height of approximately \SI{134}{m} that is in agreement with the observed upper limit of the thickness of the rings of Saturn \citep{lane-etal-1982}.
Furthermore, planetary rings in our Solar System show various optical depths.
\citet{cuzzi-etal-1980} found a maximum optical depth in the rings of Saturn of about \num{1.5} in the optical wavelength range.
\citet{esposito-etal-1983} found a trimodal distribution of optical depth in the rings of Saturn with modes at $\sim$\num{0.08}, $\sim$\num{0.5}, and $\gtrsim$\num{2.5} in the ultraviolet during stellar occultation.
In contrast, \citet{showalter-etal-1987} found very low optical depths on the order of \num{e-6} in the ring system of Jupiter.
\citet{lane-etal-1986} measured mean optical depths of $\lesssim$\num{0.6} of most of the Uranian rings, and mean optical depths in the range of \num{1.3} to \num{2.3} for the $\gamma$ and $\epsilon$ ring in the ultraviolet wavelengths.
The vertical optical depth of our ring model was set to a moderate value of \num{1} at a wavelength of \SI{0.55}{\um}.

The size of a circumplanetary ring is constrained by the Hill sphere, in which the gravitational force is dominated by the planet, and the Roche limit.
For radii larger than the Roche limit, particles potentially gather to form satellites.
The radius of the Hill sphere and the (fluid) Roche limit are given by \citep[e.g.,][]{goldreich-etal-2004, schlichting-sari-2008}%
\begin{equation}
    \label{eq:hill-radius}
    r_\mathrm{H} \approx a \left( \frac{m_\mathrm{p}}{3 M_\star} \right)^{1/3}
\end{equation}
and \citep[e.g.,][]{murray-dermott-1999, de-pater-lissauer-2001}%
\begin{equation}
    \label{eq:roche-limit}
    r_\mathrm{R} \approx \num{2.46} r_\mathrm{p} \left( \frac{\rho_\mathrm{p}}{\rho_\mathrm{r}} \right)^{1/3} \approx \num{1.53} \left( \frac{m_\mathrm{p}}{\rho_\mathrm{r}} \right)^{1/3}.
\end{equation}
In Eq.~(\ref{eq:hill-radius}), $a$ is the semimajor axis, $m_\mathrm{p}$ is the planetary and $M_\star$ is the stellar mass, and in Eq.~(\ref{eq:roche-limit}), $\rho_\mathrm{p}$ and $\rho_\mathrm{r}$ are the density of the planet and its ring, respectively.
Since $r_\mathrm{H} \propto a$, $r_\mathrm{R} \propto \rho_\mathrm{r}^{-1/3}$, and rings of close-in extrasolar planets probably consist of rocky material (see Eq.~\ref{eq:ice-distance}), the probability of ring formation decreases with decreasing semimajor axis.
For example, a close-in Saturn-like planet with a semimajor axis of \SI{0.1}{au}, the Hill radius is approximately \num{11}$r_\mathrm{p}$.
With a ring density of \SI{3.5}{g.cm^{-3}}, assuming rocky, that is, dense silicate material, the Roche limit is about \num{1.4}$r_\mathrm{p}$.
However, \citet{schlichting-chang-2011} found that most of the currently known extrasolar planets have the potential to harbor circumplanetary rings.
In addition, the lifetime of close-in extrasolar rings are strongly limited by the radiation drag or Poynting-Robertson force \citep{poynting-1904, robertson-1937}.
The decay time of a ring particle in an optically thin ring is proportional to its size \citep{burns-etal-1979, goldreich-tremaine-1982}, while for an optically thick ring, the decay timescales with the surface density of the disk \citep{schlichting-chang-2011}.
Thus, smaller particles can survive longer in an optically thick ring, with lifetimes ranging from \SIrange{e6}{e9}{years} \citep{schlichting-chang-2011}.

\subsection{Single scattering properties}
\label{subsec:single-scattering}

\begin{figure}
    \centering
    \includegraphics[width=8.8cm]{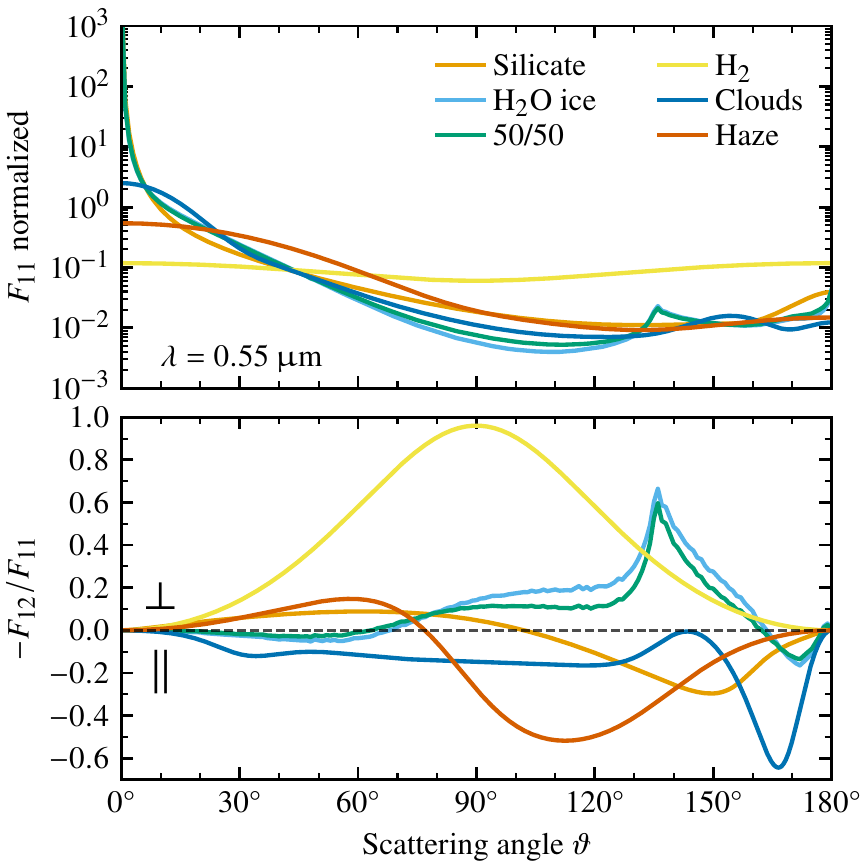}
    \caption{Normalized scattering matrix element $F_{11}$ and ratio $-F_{12} / F_{11}$ as a function of the scattering angle, $\vartheta$, for the considered chemical compositions of the ring and atmospheric particles at a wavelength of \SI{0.55}{\um}.
    Forward scattering is at $\vartheta = \ang{0}$.
    For single scattered radiation, it is $\alpha = \ang{180} - \vartheta$.
    See Sect.~\ref{subsec:single-scattering} for details.}
    \label{fig:scatmat}
\end{figure}

The ratio $-F_{12} / F_{11}$ describes the degree and orientation (see Eq.~\ref{eq:pol-angle}) of polarization for incoming unpolarized radiation that is scattered once at a single particle.
For $-F_{12} / F_{11} < 0$, the radiation is polarized parallel to the scattering plane.
Figure~\ref{fig:scatmat} shows the normalized matrix element $F_{11}$ and the ratio $-F_{12}/F_{11}$ of the scattering matrix (see Eq.~\ref{eq:stokes-scattering}) as a function of the scattering angle $\vartheta$ at a wavelength of \SI{0.55}{\um} for various considered particles.
These include molecular hydrogen, Jupiter-like haze and cloud particles \citep{mclean-etal-2017}, silicate particles with refractive indices from \citep{draine-2003}, water ice particles with refractive indices from \citep{warren-brandt-2008}, and particles consisting of a mixture of \SI{50}{\percent} silicate and \SI{50}{\percent} water ice.
For the haze and cloud particles, the size distribution described by Eq.~(\ref{eq:size-clouds}) with $s_\mathrm{eff} = \SI{0.2}{\um}$ and $\varv_\mathrm{eff} = \num{0.01}$ as well as $s_\mathrm{eff} = \SI{0.5}{\um}$ and $\varv_\mathrm{eff} = \num{0.05}$ was used, respectively.
For the ring particles, the size distribution described by Eq.~(\ref{eq:size-ring}) with $q = \num{-3}$ was used.

Since the matrix elements are a function of the particle size, they were averaged over the size distribution as follows:
\begin{equation}
    F_{ij} = \frac{ \int n(s) F_{ij}(s)\, \mathrm{d} s }{ \int n(s)\, \mathrm{d} s }.
\end{equation}
Furthermore, the element $F_{11}$ was normalized, such that $\oint F_{11}\, \mathrm{d} \Omega = 1$.
Both the silicate and the water ice particles, as well as the mixture of both, show strong forward scattering.
In addition, particles containing water ice show a maximum in the linear polarization at a scattering angle of about \ang{140}.
However, this polarization feature may be absent for nonspherical particles (see Sect.~\ref{subsec:ring-model}).
Particles consisting of silicates have a polarization maximum at a scattering angle of about \ang{150} with an opposite sign.
For silicate particles, the radiation is polarized parallel to the scattering plane at scattering angles $\gtrsim$\ang{100}.
For icy material, it is polarized parallel for scattering angles $\lesssim$\ang{70} and $\gtrsim$\ang{160}.


\section{Results}
\label{sec:results}

\subsection{Impact of planetary rings on the net polarization: General remarks}
\label{subsec:results-general}

\begin{figure*}
    \centering
    \includegraphics[width=18cm]{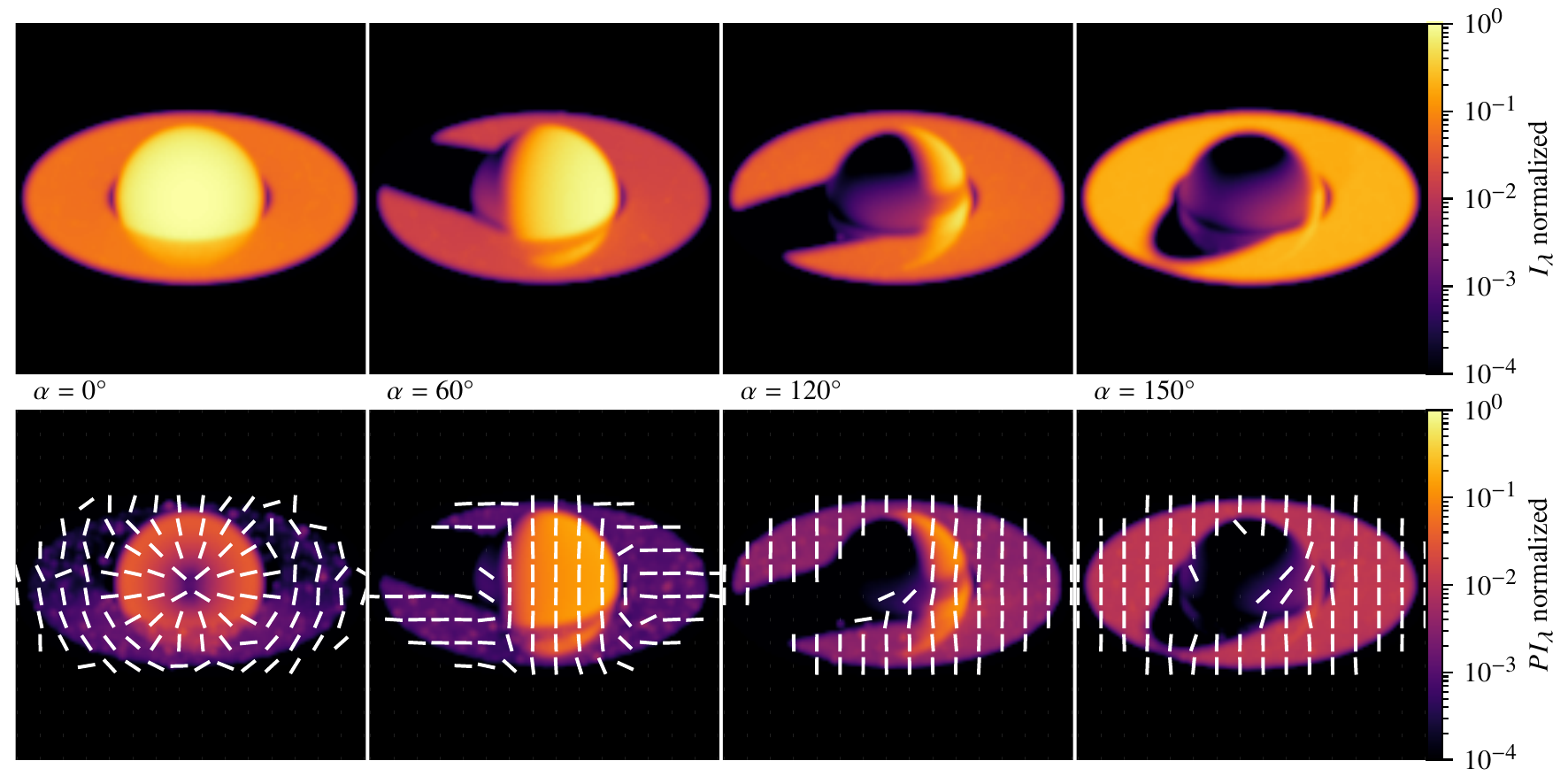}
    \caption{Total reflected flux, $I_\lambda$ (\emph{top}), and polarized flux, $PI_\lambda$ (\emph{bottom}), at four different planetary phase angles, $\alpha$, at a wavelength of \SI{0.55}{\um}.
    The fluxes are normalized to the maximum total flux at a phase angle of \ang{0}.
    Here, it is $i = \ang{30}$ and $i_\mathrm{r} = \ang{60}$ (see Fig.~\ref{fig:setup}).
    See Sect.~\ref{subsec:results-general} for details.}
    \label{fig:resolved}
\end{figure*}

\begin{figure}
    \centering
    \includegraphics[width=8.8cm]{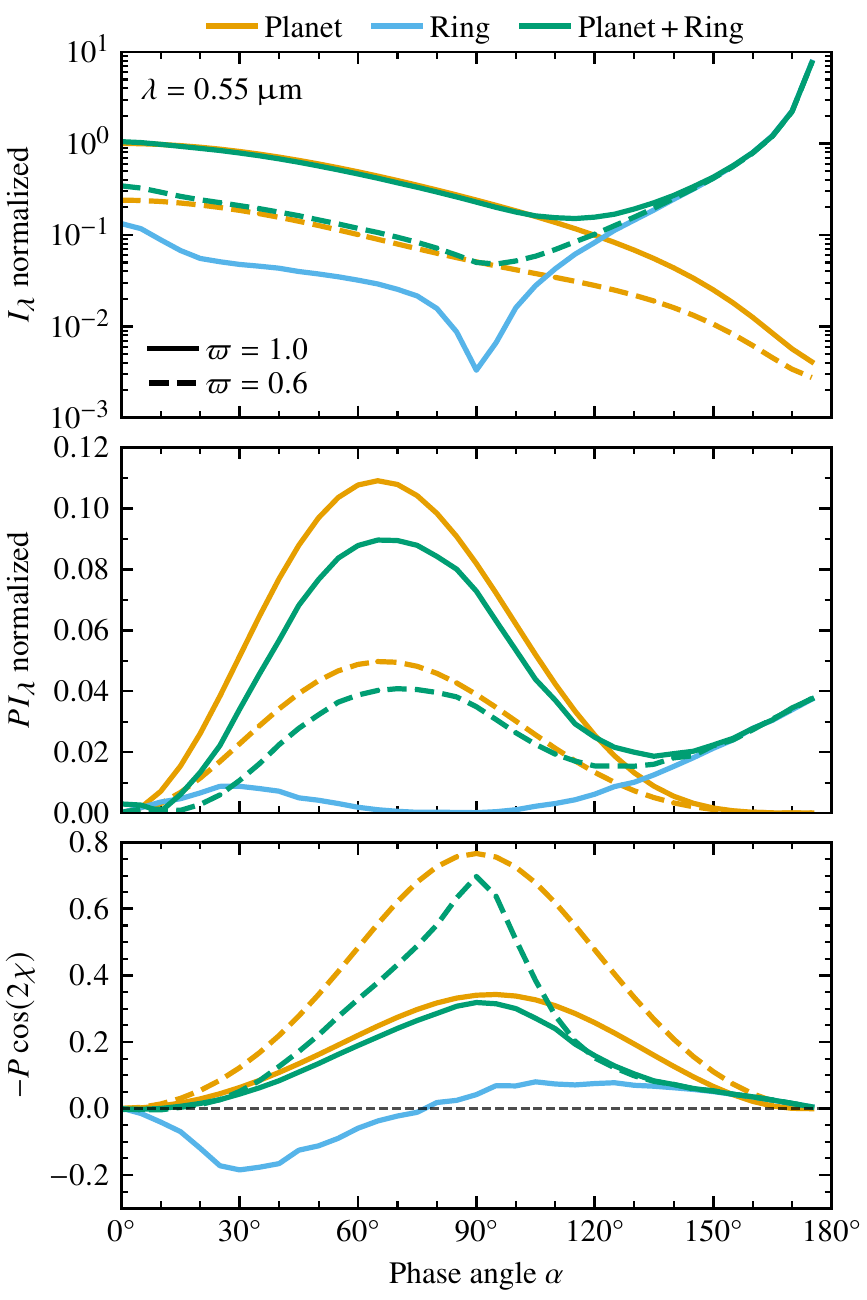}
    \caption{Total reflected flux, $I_\lambda$ (\emph{top}), polarized flux, $PI_\lambda$ (\emph{middle}), and degree of polarization, $P$ (\emph{bottom}), as a function of the planetary phase angle, $\alpha$, of a planet, a circumplanetary ring, and the combination of the two.
    The fluxes are normalized to the total reflected flux of a ringless planet at a phase angle of \ang{0} at a wavelength of \SI{0.55}{\um}.
    The line styles correspond to different single scattering albedos, $\varpi$, of the gaseous particles (except for the ring model).
    For a single scattering albedo of \num{1.0} and {0.6}, the planet has a geometric albedo of about \num{0.72} and \num{0.17}, respectively.
    See Sect.~\ref{subsec:results-general} for details.}
    \label{fig:planet-ring}
\end{figure}

First, we investigated the impact of the ring geometry.
For this purpose, we considered: (i) a planet with a radius of \SI{7e7}{m} and a simple cloud-free Rayleigh scattering atmosphere, (ii) a single ring with an obliquity or inclination of \ang{30} without a planet, and (iii) a ringed planet, which is case 1 and 2 combined.

Figure~\ref{fig:resolved} shows the spatially resolved total reflected and polarized flux maps of the planet with a circumplanetary ring at a wavelength of \SI{0.55}{\um}.
The fluxes are normalized to the maximum total flux at a phase angle of \ang{0}.
Depending on the planetary phase angle, shadows occur on the circumplanetary disk or the planet, decreasing the total reflected and polarized flux.
Furthermore, the lower part of the planet is covered by the ring, decreasing the total reflected and polarized flux of the planet as well.

For a planetary phase angle of \ang{0}, the orientation of linear polarization is pointing in radial direction toward the planet.
For increasing phase angles, the orientation of the linear polarization flips.
At a phase angle of \ang{60}, the orientation of the linear polarization of the disk is parallel with respect to the scattering plane, while for the planet, it is perpendicular.
The orientation of linear polarization of the circumplanetary ring flips, being oriented perpendicular to the scattering plane at larger phase angles of \ang{120} or \ang{150}.
At these phase angles, the observed flux is dominated by radiation scattered by the dust particles in the ring.
In addition, the stellar radiation that is reflected by the ring onto the night side of the planet (ringshine) is also visible.
In contrast, stellar radiation that is reflected by the planetary atmosphere onto the ring (planetshine) is not visible because the directly scattered light dominates the net flux.

Figure~\ref{fig:planet-ring} shows the total reflected flux, polarized flux, and the degree of polarization as a function of the planetary phase angle of the three models.
The step width of the phase angle is \ang{5}.
The fluxes are normalized to the total reflected flux of a ringless planet at a phase angle of \ang{0} at a wavelength of \SI{0.55}{\um}.
Following \citet{chernova-etal-1993}, the degree of polarization is multiplied by $(-\cos2\chi)$.
Thus, the sign of the resulting polarization describes the orientation of polarization, so that if the resulting polarization value is positive or negative, the orientation is perpendicular or parallel to the scattering plane, respectively.
In addition, for the degree of polarization, only the scattered radiation but not the unpolarized stellar flux is considered.

At a wavelength of \SI{0.55}{\um}, the planet has a somewhat high geometric albedo of about \num{0.72}, because of the non-absorbing atmosphere.
Consequently, the polarized flux of the planet provides an upper limit and is expected to be lower for hot Jupiters (see Sect.~\ref{subsec:atmospheric-model}), so the contrast of the planet and the ring increases.
For this reason, a planetary atmosphere with a single scattering albedo of \num{0.6} for H$_2$, resulting in a geometric albedo of about \num{0.17}, is also shown in Fig.~\ref{fig:planet-ring}.

If no circumplanetary ring is present, the total reflected flux is largest at a phase angle of \ang{0}, and decreases with increasing phase angle.
The reflected polarized flux has its maximum at a phase angle of about \ang{65} and decreases with increasing as well as decreasing phase angle.
The degree of polarization has a maximum value of about \num{0.33}, centered around a phase angle of about \ang{95}.
In addition, for a lower single scattering albedo, the reflected flux decreases whereas the degree of polarization increases due to the absorption and reduced multiple scattering.
Along the planetary orbit, the radiation is polarized perpendicular to the scattering plane.
These results are consistent with previous studies by, for example, \citet{buenzli-schmid-2009} or \citet{bailey-etal-2018}.

For the second model, solely a ring with an inclination of \ang{30} was considered.
Here, the total reflected and polarized flux are usually below the total and polarized planetary flux.
First, at phase angles of $\gtrsim$\ang{120}, the total reflected flux of the ring exceeds the total reflected planetary flux.
This is due to the small dust particles that preferentially scatter the radiation in the forward direction.
At phase angles of $\gtrsim$\ang{130}, the polarized flux of the ring dominates the net polarization.
The total reflected flux decreases at a phase angle of \ang{90} because the surface area being illuminated is smallest, that is, at the equinox of the planet where the ring is facing the star edge-on.
The degree of polarization is largest at a phase angle of about \ang{30} with a value of about \num{0.2}, and decreases to \num{0} at phase angles of $\sim$\ang{85}.
For smaller phase angles, the radiation is polarized parallel to the scattering plane, while it is polarized perpendicular to the scattering plane, for larger planetary phase angles.

Finally, we considered the combination of the previous models, that is, a planet with an additional circumplanetary ring.
For a purely scattering atmosphere, the behavior of the total reflected flux is comparable to the flux of a planet without a ring at phase angles $\lesssim$\ang{110}.
In contrast, for an absorbing atmosphere, the total flux shows a significant contribution of the ring at small phase angles as well.
Although, the total reflected flux strongly increase due to scattering by the ring particles at large phase angles, shadows on the circumplanetary ring and on the planet are slightly decreasing the total flux.
The polarized flux of the planet with a ring is somewhat lower compared to that of the planet without a ring because the polarized flux of the ring particles is small around a phase angle of \ang{90} and it is polarized parallel to the scattering plane for phase angles $\lesssim$\ang{90}.
The degree of polarization is largest at a phase angle of about \ang{90}.
In contrast to a planet without a ring, the polarization distribution of the ringed planet shows a skewness.
In particular, at phase angles $>$\ang{90}, the degree of polarization decreases stronger compared to the degree of polarization of a planet without a ring.
The impact of this effect increases if the single scattering albedo of the atmospheric particles decreases.
Finally, at phase angles of $\gtrsim$\ang{140}, the behavior of the degree of polarization mirrors the single scattering properties of the individual particles, that is, $-F_{12} / F_{11}$, at scattering angles $\lesssim$\ang{40} (see Fig.~\ref{fig:scatmat}).

The increase in the scattered flux at large phase angles ($\gtrsim$\ang{120}) applies also for the total reflected flux of extrasolar planets covered by oceans \citep{williams-gaidos-2008, robinson-etal-2010, zugger-etal-2010, zugger-etal-2011-erratum, trees-stam-2019}.
However, the total reflected flux of the Sun glint is wavelength-dependent due to the atmosphere above.
In particular, the Sun glint disappears in the total reflected flux at short wavelengths ($\sim$\SI{0.4}{\um}) where the atmosphere becomes thicker, whereas the strength of the glint increases in the total reflected flux with increasing wavelength.
In contrast, the increased flux of the ring only depends on the ring size and its particle properties.

\subsection{Impact of the ring size}
\label{subsec:results-ring-size}

\begin{figure}
    \centering
    \includegraphics[width=8.8cm]{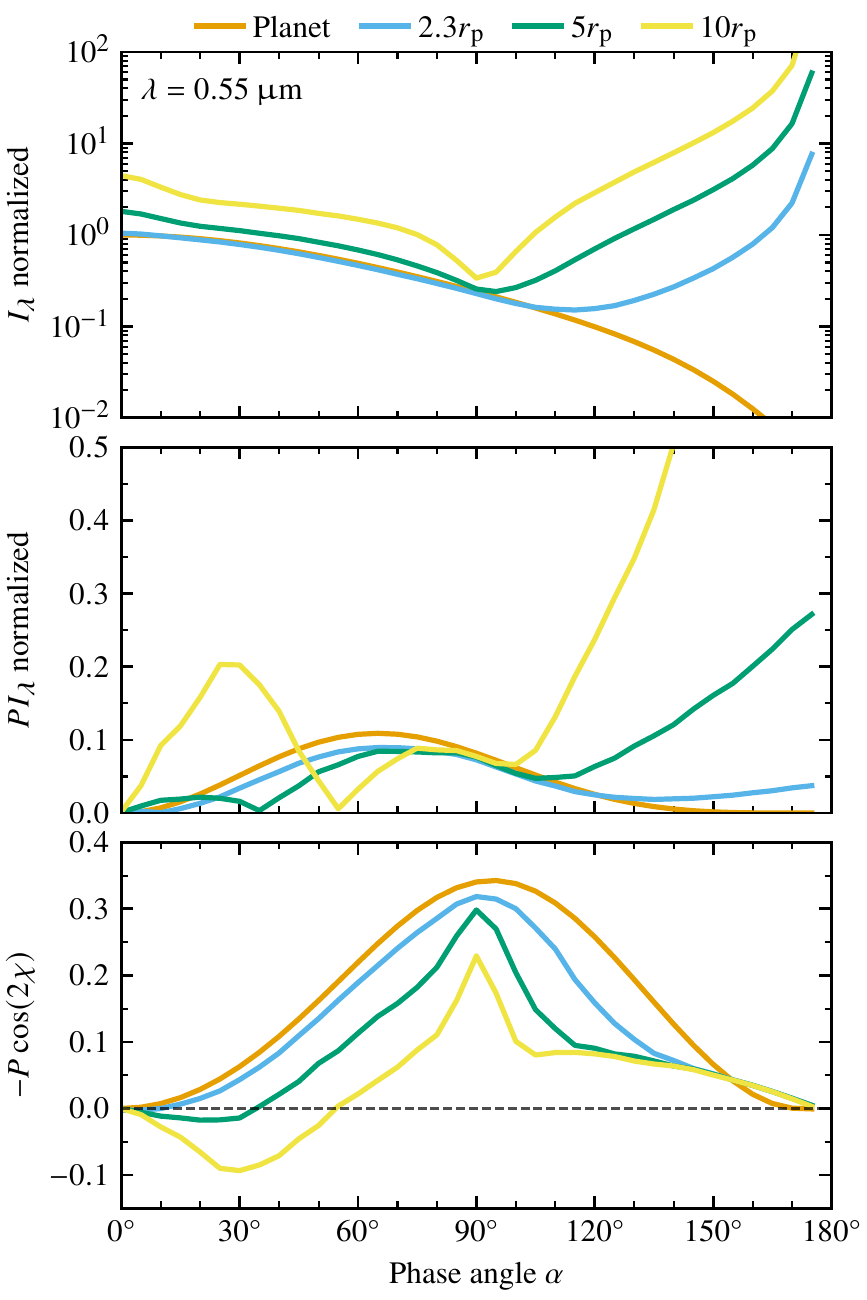}
    \caption{Total reflected flux, $I_\lambda$ (\emph{top}), polarized flux, $PI_\lambda$ (\emph{middle}), and degree of polarization, $P$ (\emph{bottom}), as a function of the planetary phase angle, $\alpha$, of a circumplanetary ring with different outer radii.
    For comparison, the flux of a planet without a ring is also shown.
    The fluxes are normalized to the total reflected flux of a ringless planet at a phase angle of \ang{0} at a wavelength of \SI{0.55}{\um}.
    See Sect.~\ref{subsec:results-ring-size} for details.}
    \label{fig:ring-size}
\end{figure}

As shown in the previous section, a circumplanetary ring with a radius of \num{2.3}$r_\mathrm{p}$ has a significant impact on the total reflected and polarized radiation at large phase angles.
Since the total reflected radiation depends on the size of the circumplanetary ring, we compared rings with different outer radii of \num{2.3}$r_\mathrm{p}$, \num{5}$r_\mathrm{p}$, and \num{10}$r_\mathrm{p}$.
Figure~\ref{fig:ring-size} shows the total reflected flux, polarized flux, and the degree of polarization as a function of the planetary phase angle for the different considered outer radii of the ring at a wavelength of \SI{0.55}{\um}.
For comparison, the total reflected flux of a planet without a disk is shown as well.

With increasing ring radius, the total flux increases as well.
In particular, at a phase angle of \ang{0}, for a ring radius of \num{5}$r_\mathrm{p}$ and \num{10}$r_\mathrm{p}$, the flux is about \num{1.7} and \num{4} times larger, respectively.
Especially at large phase angles of $\gtrsim$\ang{150}, the total reflected flux increases by a factor of $\sim$\num{10}.
In addition, the polarized radiation increases as well, and the polarized flux is dominated by radiation scattered by dust in the circumplanetary ring.
Subsequently, the net radiation is polarized parallel to the scattering plane at phase angles $\lesssim$\ang{35} and $\lesssim$\ang{60} for a ring radius of \num{5}$r_\mathrm{p}$ and \num{10}$r_\mathrm{p}$, respectively.
For an outer ring radius of \num{10}$r_\mathrm{p}$, the degree of polarization has a local maximum at a phase angle of \ang{30}, which can also be seen in the single scattering properties of the silicate particles at a scattering angle of \ang{150} (see Fig.~\ref{fig:scatmat}).
In contrast to a planet without a circumplanetary ring, the distribution of the polarization degree becomes sharper at a phase angle of \ang{90}.
In addition, there is zero polarization and a change in the orientation of polarization at phase angles of about \ang{35} or \ang{60}, depending on the ring radius.

\subsection{Impact of the ring composition}
\label{subsec:results-composition}

\begin{figure}
    \centering
    \includegraphics[width=8.8cm]{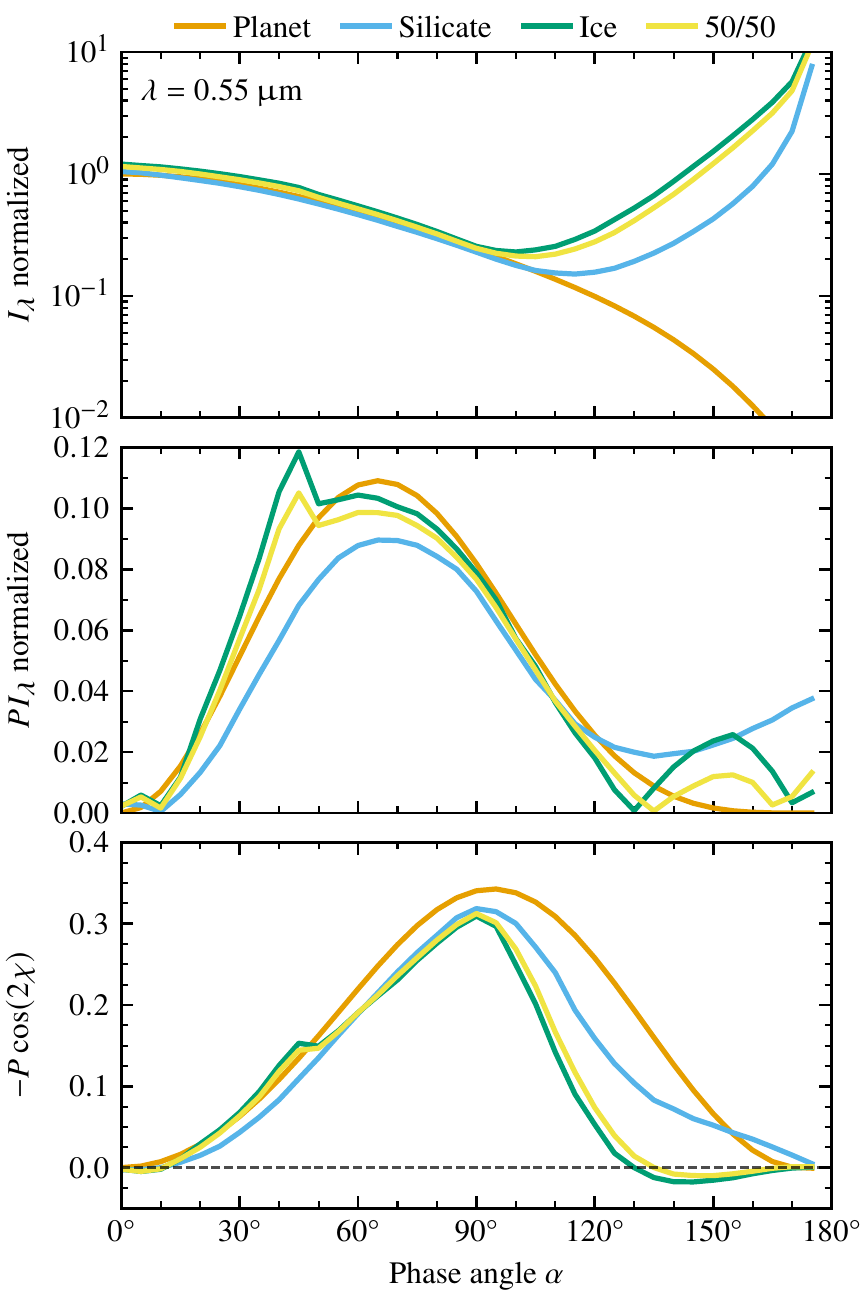}
    \caption{Same as Fig.~\ref{fig:ring-size}, but for various ring compositions.
    See Sect.~\ref{subsec:results-composition} for details.}
    \label{fig:ring-composition}
\end{figure}

Besides geometrical and orbit parameters, the total reflected and polarized flux also depends on the optical properties of the dust particles inside the ring.
In this section we investigated the impact of the ring composition.
For this purpose, we also assumed a ring composed of water ice particles, and a mixture of \SI{50}{\percent} silicate and \SI{50}{\percent} water ice.

Figure~\ref{fig:ring-composition} shows the total reflected flux, polarized flux, and the degree of polarization as a function of the planetary phase angle for different ring compositions at a wavelength of \SI{0.55}{\um}.
For the circumplanetary ring containing icy material, the total reflected flux is somewhat larger compared to a ring containing silicate and rocky material.
In particular, at phase angles $\gtrsim$\ang{110}, the total reflected flux increases strongly if icy material is present in the circumplanetary ring.
Furthermore, at a phase angles of about \ang{135}, the polarization becomes zero and changes its orientation; thus, for phase angles $\gtrsim$\ang{135}, the radiation is polarized parallel to the scattering plane for an icy ring.
This is because of the opposite sign of the term $-F_{12} / F_{11}$ of water ice compared to silicate grains (see Fig.~\ref{fig:scatmat}).
While the polarized radiation increases for a silicate ring at phase angles $\gtrsim$\ang{140}, the polarized radiation has a local maximum at a phase angle of about \ang{155} if there are ice particles in the ring.
Finally, for icy particles, a small maximum at a phase angle of \ang{40} is visible in the polarized flux as well as in the degree of polarization.
However, this polarization feature may be absent for nonspherical particles (see Sect.~\ref{subsec:ring-model}).

\subsection{Impact of the grain size}
\label{subsec:results-particle-size}

\begin{figure}
    \centering
    \includegraphics[width=8.8cm]{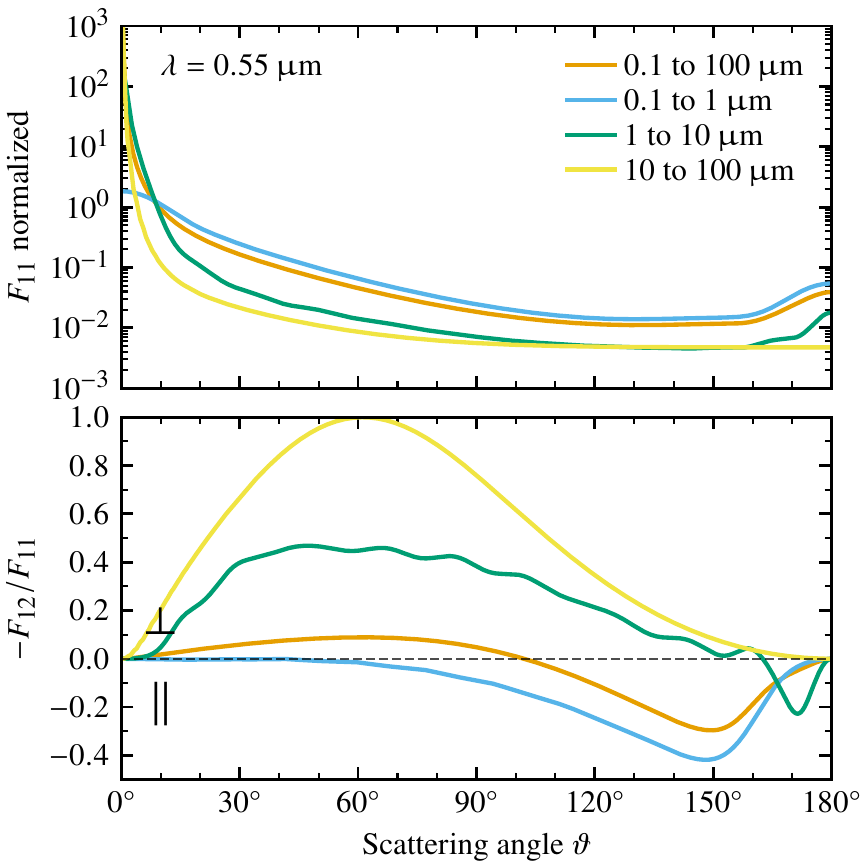}
    \caption{Same as Fig.~\ref{fig:scatmat}, but for various grain size distributions of the ring particles.
    See Sect.~\ref{subsec:results-particle-size} for details.}
    \label{fig:scatmat2}
\end{figure}

\begin{figure}
    \centering
    \includegraphics[width=8.8cm]{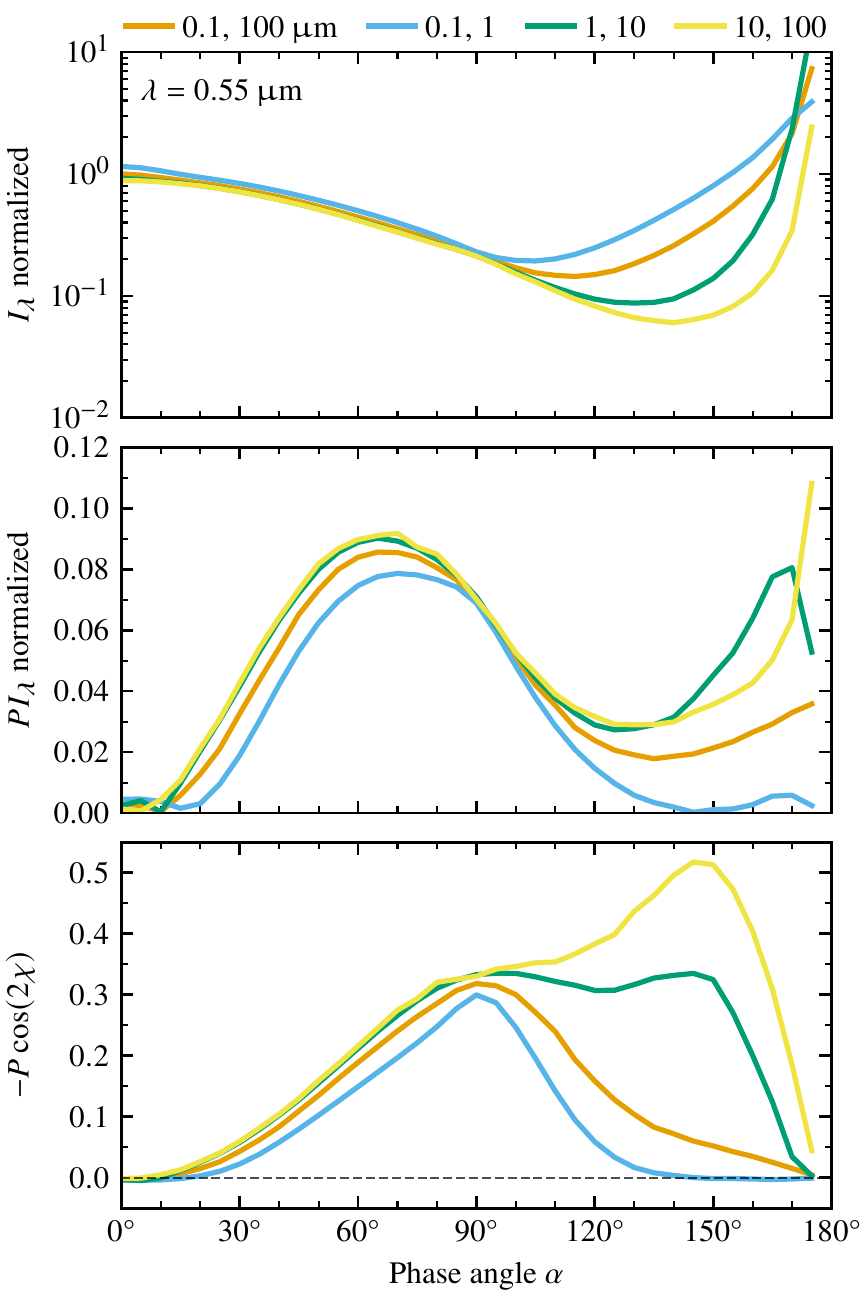}
    \caption{Same as Fig.~\ref{fig:ring-size}, but for various grain sizes.
    See Sect.~\ref{subsec:results-particle-size} for details.}
    \label{fig:particle-size}
\end{figure}

The scattering-angle-dependent reflected radiation and polarization degree strongly depend on the particle size, more specifically on the size parameter $x = 2\pi s / \lambda$.
To investigate the impact of the particle size, we assumed the same size distribution as defined by Eq.~(\ref{eq:size-ring}) with $q = -3$, but varied the minimum and maximum grain size.
Figure~\ref{fig:scatmat2}  shows the normalized matrix element $F_{11}$ and the ratio $-F_{12}/F_{11}$ of the scattering matrix as a function of the scattering angle $\vartheta$ at a wavelength of \SI{0.55}{\um} for various grain size distributions of the ring particles.
In addition to the previous broad size distribution with grain sizes ranging from \SIrange{0.1}{1}{\um}, we considered narrow size distributions ranging from \SIrange{0.1}{1}{\um}, \SIrange{1}{10}{\um}, and \SIrange{10}{100}{\um}.
For increasing particles sizes, the magnitude of forward scattering increases, whereas back scattering decreases.
Furthermore, the degree of polarization strongly increases at a scattering angle of \ang{60} up to $\sim$\SI{100}{\percent} for the largest considered particles.
For the smallest considered particles, the degree of polarization has a maximum of $\sim$\SI{42}{\percent} at a scattering angle of about \ang{150}, but the radiation is polarized parallel to the scattering plane.

Figure~\ref{fig:particle-size} shows the total reflected flux, polarized flux, and the degree of polarization as a function of the planetary phase angle for various grain sizes.
The distribution with the smallest particles (\SIrange{0.1}{1}{\um}) shows the highest reflected flux at all phase angles.
In contrast, the polarized flux is the lowest compared to the other grain sizes.
As a result, the degree of polarization is small as well.
In particular, at phase angles of $\gtrsim$\ang{140}, the degree of polarization is almost zero.
For increasing grain size, the reflected flux decreases, especially at large phase angles ($\sim$\ang{150}).
However, the polarized flux increases, resulting in a maximum degree of polarization at a phase angle of about \ang{150}.
Since the size distribution strongly decreases with increasing grain size ($q = -3$), these features of large particles are lost in the broad distribution (see Fig.~\ref{fig:scatmat2}).

\subsection{Impact of the ring inclination and orientation}
\label{subsec:results-inclination}

\begin{figure*}
    \centering
    \includegraphics[width=18cm]{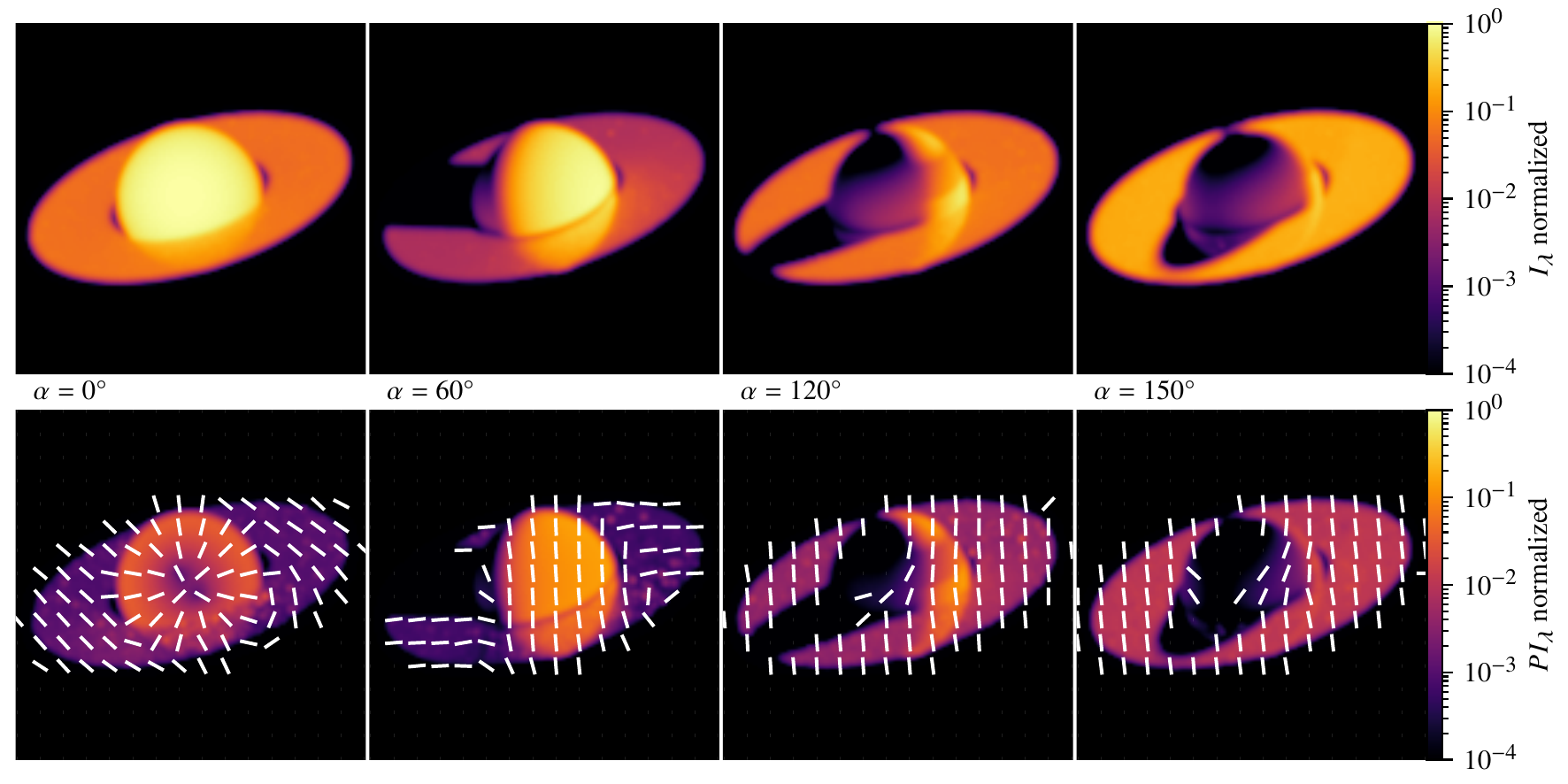}
    \caption{Same as Fig.~\ref{fig:resolved}, but for a ring orientation of \ang{30} (instead of \ang{0}), resulting in an angle of $i_\mathrm{r} \approx \ang{64.3}$ between the observer and ring normal (see Fig.~\ref{fig:setup}).
    See Sect.~\ref{subsec:results-inclination} for details.}
    \label{fig:resolved-orientation}
\end{figure*}

\begin{figure}
    \centering
    \includegraphics[width=8.8cm]{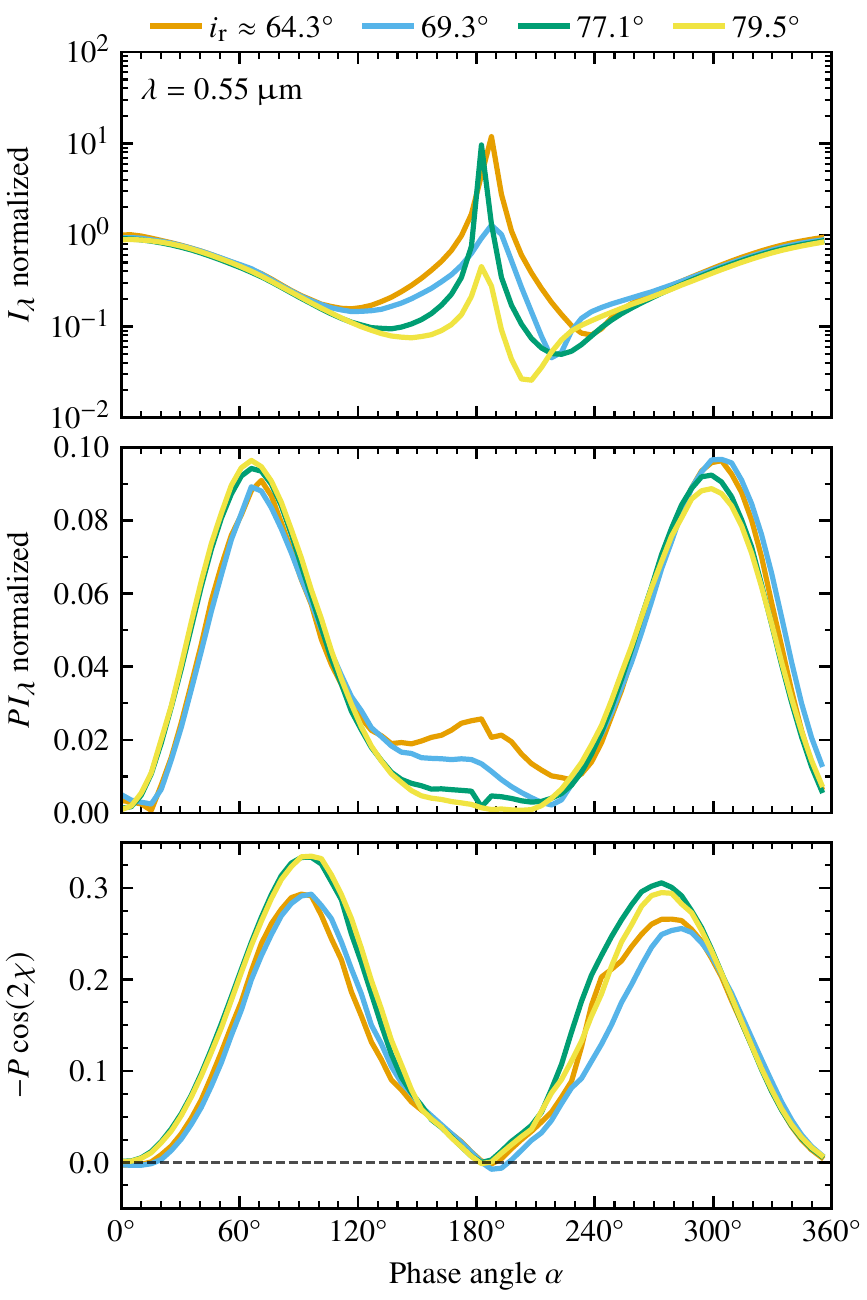}
    \caption{Same as Fig.~\ref{fig:ring-size}, but for various ring inclinations.
    See Sect.~\ref{subsec:results-inclination} for details.}
    \label{fig:ring-inclination}
\end{figure}

Figure~\ref{fig:resolved-orientation} shows the spatially resolved total reflected and polarized flux for different phase angles, with a ring inclination of \ang{30}.
In contrast to Fig.~\ref{fig:resolved}, the longitude or orientation of the ring is rotated by an angle of \ang{30} (an orientation of \ang{90} corresponds to an edge-on view).
Thus, the angle, $i_\mathrm{r}$, between observer and ring normal amounts to $\sim$\ang{64.3} (see Fig.~\ref{fig:setup}).
However, for single scattered radiation, and thus the fraction with the typically highest contribution to the net scattered flux, the scattering angle is still $\vartheta = \ang{180} - \alpha$.
Thus, at a phase angle of \ang{60}, the radiation scattered in the ring has an orientation of linear polarization parallel to the scattering plane, while it is perpendicular for the presented cases for phase angles \ang{120} and \ang{150}.
Although, the scattering angle for single scattered radiation does not change, varying the inclination and orientation of the ring alters the observable ring area, and the observed shadowing on the planet and its ring.
Similar to Fig.~\ref{fig:resolved}, the night side of the planet is partially illuminated by ringshine.

Figure~\ref{fig:ring-inclination} shows the total reflected flux, polarized flux, and the degree of polarization as a function of the planetary phase angle for various ring inclinations.
Here, the results are shown for an entire planetary orbit, that is, for phase angles from \ang{0} to \ang{360}.
For $i_\mathrm{r} = \ang{64.3}$ and \ang{69.3}, the ring inclination is $i = \ang{30}$, and the orientation is rotated by \ang{30} and \ang{45}, respectively.
For $i_\mathrm{r} = \ang{77.1}$ and \ang{79.5}, the ring inclination is $i = \ang{15}$, and the orientation is rotated by \ang{30} and \ang{45}, respectively.

The total reflected flux has a maximum at a phase angle of about \ang{180} due to the forward scattering particles in the circumplanetary ring.
In addition, depending on the orientation of the ring, there is a minimum in the total flux at phase angles between \SIrange{200}{240}{\degree}, because the illuminated surface of the ring decreases strongly at these angles, that is, at the equinox of the planet.
This minimum is only at phase angles greater than \ang{180}, because at these phase angles, the ring is facing edge-on the star with a small planetary contribution at the same time.
The polarized flux shows maxima at a phase angle of about \ang{70} and, depending on the orientation, at \SIrange{290}{300}{\degree}.
The maximum degree of polarization at these phase angles also varies with the ring orientation.
In addition, at phase angles of about \ang{180}, the polarized flux is largest for $i_\mathrm{r} \approx \ang{64.3}$, whereas it decreases for increasing $i_\mathrm{r}$.

\subsection{Spectropolarimetric signatures}
\label{subsec:results-cloudy}

\begin{figure}
    \centering
    \includegraphics[width=8.8cm]{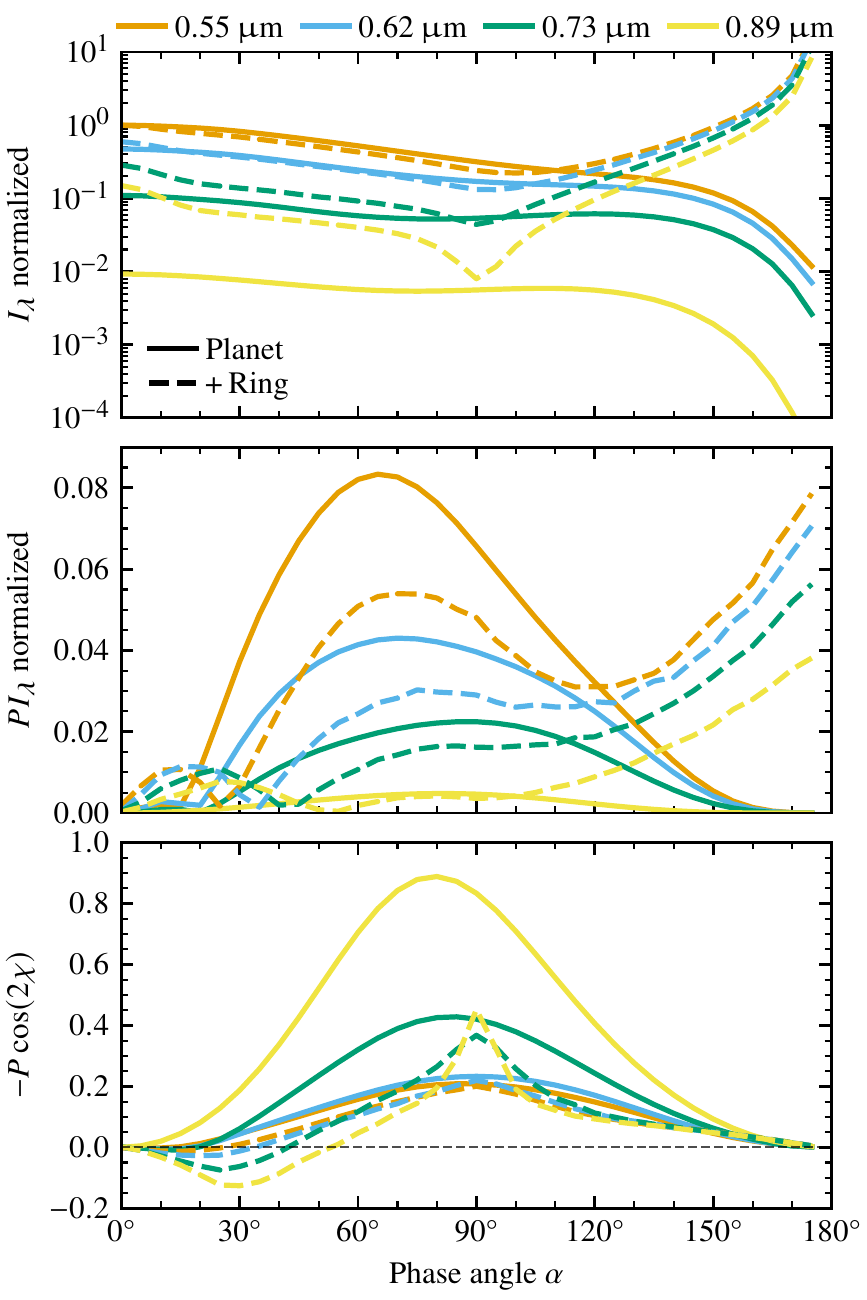}
    \caption{Same as Fig.~\ref{fig:ring-size}, but for a cloudy planet at various wavelengths.
    At a wavelength of \SI{0.55}{\um}, \SI{0.62}{\um}, \SI{0.73}{\um}, and \SI{0.89}{\um}, the planet has a geometric albedo of about \num{0.27}, \num{0.14}, \num{0.04}, and \num{0.005}, respectively.
    See Sect.~\ref{subsec:results-cloudy} for details.}
    \label{fig:cloudy-phase}
\end{figure}

\begin{figure}
    \centering
    \includegraphics[width=8.8cm]{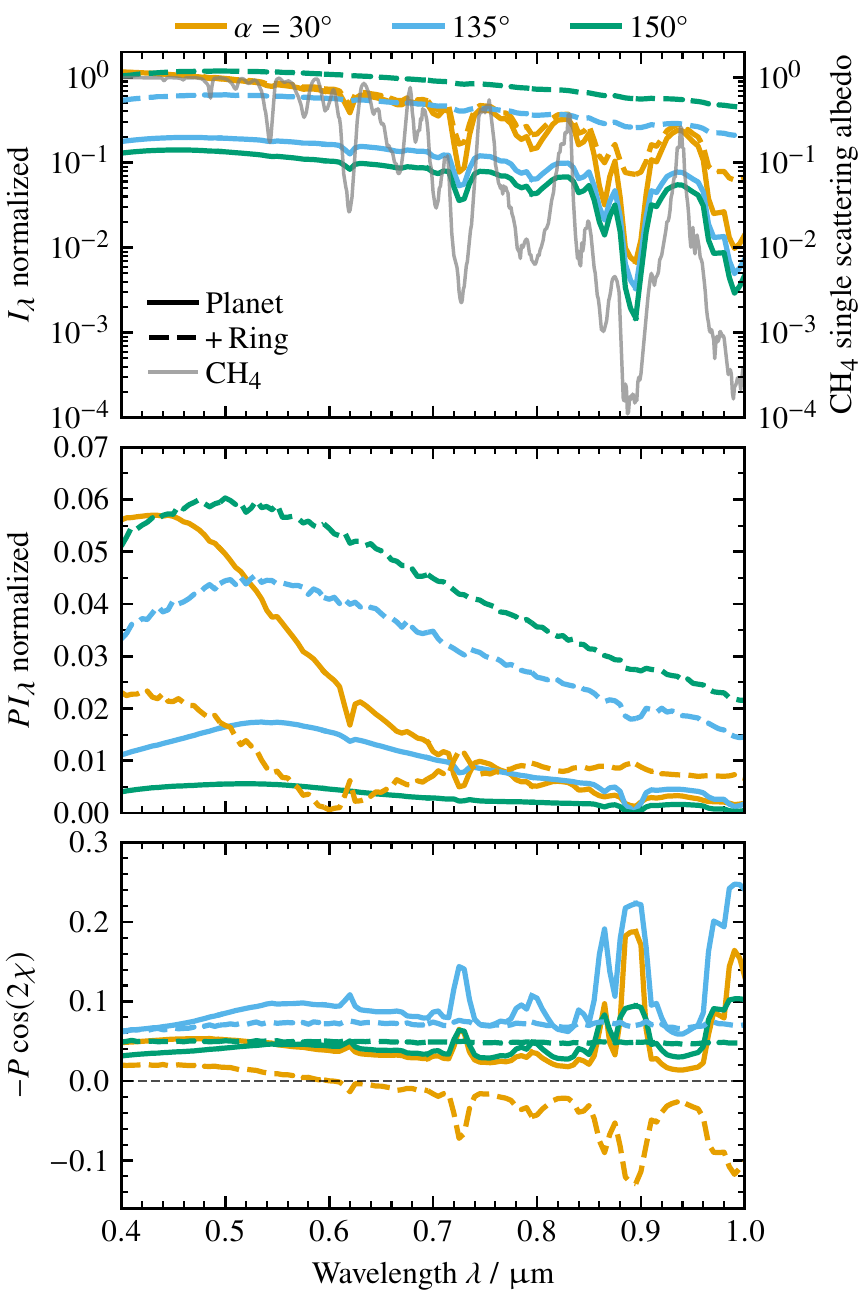}
    \caption{Total reflected flux, $I_\lambda$ (\emph{top}), polarized flux, $PI_\lambda$ (\emph{middle}), and degree of polarization, $P$ (\emph{bottom}), as a function of wavelength, $\lambda$, for a cloudy atmosphere.
    The fluxes are normalized to the total reflected flux of a ringless planet at a phase angle of \ang{0} at a wavelength of \SI{0.55}{\um}.
    At this wavelength, the planet has a geometric albedo of about \num{0.27}.
    In addition, the single scattering albedo of CH$_4$ is shown in the top panel.
    See Sect.~\ref{subsec:results-cloudy} for details.}
    \label{fig:cloudy-spectra}
\end{figure}

In this section we investigated the wavelength dependence of the total reflected and polarized flux.
For this purpose, we assumed that methane is present in the atmosphere to account for a wavelength-dependent absorption.
The absorption cross section of methane was taken from \citet{karkoschka-1994} with a volume mixing ratio of \num{1.8e-3}, similar to what was found in Jupiter \citep{sato-hansen-1979}.
The refractive index and the scattering cross section of methane were calculated using the formulas given by \citet{sneep-ubachs-2005}.
Furthermore, cloud and haze particles are present in the atmosphere to represent the atmospheric structure and composition of Jupiter (see Sect.~\ref{subsec:atmospheric-model}).
The cloud and haze properties are listed in Table~\ref{tab:model-parameter}.

Figure~\ref{fig:cloudy-phase} shows the total reflected flux, polarized flux, and the degree of polarization as a function of the planetary phase angle at various wavelengths for a cloudy planet with and without a circumplanetary ring containing silicate particles.
There are methane bands at the considered wavelengths of \SI{0.62}{\um}, \SI{0.73}{\um}, and \SI{0.89}{\um}.
Since the strength of the absorption increases with increasing wavelength, the total reflected flux decreases with increasing wavelength.
At a wavelength of \SI{0.55}{\um}, \SI{0.62}{\um}, \SI{0.73}{\um}, and \SI{0.89}{\um}, the planet has a geometric albedo of about \num{0.27}, \num{0.14}, \num{0.04}, and \num{0.005}.
However, with increasing wavelength, the total reflected flux is dominated by the scattered radiation of the circumplanetary ring, because most radiation entering the atmosphere is absorbed.
In particular, at a wavelength of \SI{0.73}{\um} and \SI{0.89}{\um}, the reflected flux is dominated by radiation scattered in the ring.

The degree of polarization is largest for the planet without a ring at a wavelength of \SI{0.89}{\um}, because most of the radiation is absorbed, and only radiation scattered by the top gaseous layers with a high degree of polarization reaches the observer.
For a planet with a circumplanetary ring, the distribution of the degree of polarization at a wavelength of \SI{0.73}{\um} and \SI{0.89}{\um} is about similar to the case of an increased ring radius (see Sect.~\ref{subsec:results-ring-size}).
However, the maximum value at a phase angle of \ang{90} is somewhat larger compared to an increased ring radius.
For a wavelength of \SI{0.89}{\um}, there is a maximum in the polarization at a phase angle of about \ang{30} that is polarized parallel to the scattering plane, zero polarization at about \ang{55}, and a sharp maximum at a phase angle of \ang{90}.

Figure~\ref{fig:cloudy-spectra} shows spectra of the total reflected flux, polarized flux and degree of polarization for two models at three different planetary phase angles.
Here, we considered a planet without a ring and a planet with an additional circumplanetary ring.
The step width of the wavelength is \SI{5}{nm}.
The fluxes are normalized to the total reflected flux of a ringless planet at a phase angle of \ang{0} at a wavelength of \SI{0.55}{\um}.
In addition to the total reflected flux, the single scattering albedo of CH$_4$ is shown.

At a phase angle of \ang{30}, the scattered radiation decreases with increasing wavelength and shows characteristic methane absorption features.
The total reflected fluxes in both cases, a planet only and a ringed planet, are similar outside the methane band because they are dominated by the reflected planetary flux at this phase angle.
At a phase angle of \ang{135} or \ang{150}, the total reflected flux of the ringless planet decreases strongly.
In contrast, the total reflected flux of the planet with a ring is significant larger compared to a ringless planet, and it does not show methane absorption features in the spectrum.

For a planet without a ring, the degree of polarization increases in the methane bands, similar to what has been observed by \citet{schmid-etal-2011} or \citet{mclean-etal-2017}.
The radiation is polarized perpendicular to the scattering plane.
However, for a ringed planet, the radiation is polarized parallel to the scattering plane at wavelengths $\sim$\SI{0.6}{\um} due to the dust grains in the ring at a phase angle of \ang{30}.
In addition, at phase angles of \ang{135} or \ang{150}, no significant methane absorption features are visible in the polarization spectrum.
Thus, the degree of polarization is almost constant, with a value of about \num{0.05} and perpendicular to the scattering plane in the considered wavelength range.


\section{Conclusions}
\label{sec:conclusions}

In this study we have investigated the total reflected and polarized flux of a circumplanetary ring of an extrasolar planet and its impact on the net scattered light polarization of the planet-disk system.
The circumplanetary ring consisted of spherical micrometer-sized particles.
We compared the total reflected and polarized fluxes of a planet without a ring, a planet with a single ring, and a planet with a circumplanetary ring (see Sect.~\ref{subsec:results-general}).
Subsequently, we studied various parameters that have an impact on the reflected radiation.
They include the ring size (Sect.~\ref{subsec:results-ring-size}), the ring composition (Sect.~\ref{subsec:results-composition}), the size of the ring particles (Sect.~\ref{subsec:results-particle-size}), and the ring inclination and orientation (Sect.~\ref{subsec:results-inclination}).
Finally, we also included a cloudy planetary model with methane to account for absorption in the atmosphere (see Sect.~\ref{subsec:results-cloudy}).

The dust particles in the circumplanetary ring show strong forward scattering for the considered grain size distribution at the considered wavelengths.
As a result, for phase angles $\lesssim$\ang{120}, the impact of the circumplanetary ring on the total flux is insignificant because it is dominated by the total reflected and polarized flux of the planet.
However, for phase angles $\gtrsim$\ang{120}, the total scattered flux of the ring exceeds the total reflected planetary flux.
For an increasing outer radius of the ring, the total reflected flux of the ring shows an increasingly significant contribution at small phase angles.
In particular, the orientation of polarization is parallel to the scattering plane at small phase angles for a ring containing silicate particles.
In contrast, for a ring containing water ice particles, the orientation of polarization is parallel to the scattering plane at large phase angles.
For different ring inclinations and orientations, the total reflected and polarized flux show minima and maxima at specific phase angles.
In addition, the maximum degree of polarization varies depending on the ring orientation.
For smaller ring particles (\SIrange{0.1}{1}{\um}), the reflected flux is usually larger at all phase angles compared to larger particles.
In contrast, the degree of polarization is largest for large ring particles (\SIrange{10}{100}{\um}).

For a Jupiter-like atmosphere with methane and aerosols, a planet without a ring, or at small phase angles ($\sim$\ang{30}) in general, methane absorption features are found in the spectrum of the total reflected and polarized flux.
In contrast, for a ringed planet at large phase angles ($\sim$\ang{135} or $\sim$\ang{150}), the reflected and polarized flux does not show any absorption features and the degree of polarization is constant across the considered wavelength range.
Furthermore, at small phase angles, the orientation of polarization of a ringed planet changes for increasing wavelength due to the different orientation of polarization of the particles in the ring at a phase angle of \ang{30}.

As shown in Fig.~\ref{fig:planet-ring}, the reflected flux of the considered circumplanetary ring is usually smaller compared to the reflected flux of the planet.
In particular, for a planet with a geometrical albedo of about \num{0.17}, the reflected flux as a fraction of the total stellar flux is about \SI{94}{ppm} at a phase angle of \ang{0} and a semimajor axis of \SI{0.02}{au}.
For the circumplanetary ring, the reflected flux as a fraction of the total stellar flux is about \SI{50}{ppm} at the same distance.
In addition, the polarized flux decreases to about \SI{16}{ppm} at a phase angle of \ang{70} with an additional circumplanetary ring.
At a phase angle of \ang{150}, where the polarized flux of the ring exceeds the polarized flux of the planet, the polarized flux of the ring as a fraction of the total stellar flux is about \SI{8}{ppm}.
Thus, the polarized flux has to be measured at a parts-per-million level, which is currently at the limit of modern polarimeters, such as the High-Precision Polarimetric Instrument-2 \citep[HIPPI-2;][]{bailey-etal-2020}, and below the parts-per-million level needed for an increasing semimajor axis to detect the scattered polarized flux of a circumplanetary ring.
Consequently, only close-in planets, provided that rings exist at these distances, are suitable targets for current polarimetric instruments.

In this study, the observer was located in the plane of the planetary orbit, but reflected light polarimetry is not limited to transiting planets.
In particular, inclined or even face-on orbits are also expected to show characteristic variations in the polarized radiation due to the ring geometry.
However, this is beyond the scope of the current study.

In summary, we find unique features in the phase-angle- and wavelength-dependent total reflected and polarized flux due to scattering of the stellar radiation by dust in circumplanetary rings of extrasolar planets.
Thus, exoplanet polarimetry not only provides the means to study the planetary atmosphere and surface, but also to identify the existence and constrain the properties of exoplanetary rings.


\begin{acknowledgements}
    The authors thank the anonymous referee for useful comments and suggestions.
    This research made use of
    \href{https://ui.adsabs.harvard.edu}{NASA's Astrophysics Data System},
    \href{https://www.astropy.org}{Astropy}, a community-developed core Python package for Astronomy \citep{astropy-2013, astropy-2018},
    \href{https://matplotlib.org}{Matplotlib} \citep{hunter-2007},
    and \href{https://numpy.org}{Numpy} \citep{harris-etal-2020}.
\end{acknowledgements}


\section*{ORCID iDs}

M. Lietzow \orcidlink{0000-0001-9511-3371} \url{https://orcid.org/0000-0001-9511-3371}\\
S. Wolf \orcidlink{0000-0001-7841-3452} \url{https://orcid.org/0000-0001-7841-3452}


\bibliographystyle{aa}
\bibliography{bibliography.bib}

\end{document}